\pdfoutput=1

\documentclass[11pt]{article}

\usepackage[preprint]{acl}

\usepackage[utf8]{inputenc} 
\usepackage[T1]{fontenc}    
\usepackage{hyperref}       
\usepackage{url}            
\usepackage{booktabs}       
\usepackage{amsfonts}       
\usepackage{nicefrac}       
\usepackage{microtype}

\usepackage{smartdiagram}
\usepackage{framed}

\usepackage{wrapfig}

\usepackage{enumitem}

\usepackage{amsmath}
\usepackage{amssymb}

\usepackage{xcolor}         
\usepackage{multirow}
\usepackage{graphicx}
\usepackage{tabularx}
\usepackage{subcaption}

\usepackage{longtable}
\usepackage{makecell}

\usepackage{algorithmicx,algorithm}

\usepackage{xcolor}
\usepackage[title]{appendix}

\definecolor{formalshade}{rgb}{0.8,0.9,0.95}

\usepackage[strict]{changepage}
\definecolor{MyPink}{RGB}{255,178,178}
\definecolor{MyBlue}{RGB}{178,178,255}
\usetikzlibrary{er,positioning}

\definecolor{darkblue}{rgb}{0.0, 0.0, 0.55}
\definecolor{lightgray}{rgb}{0.929, 0.929, 0.929}

\definecolor{darkblue}{rgb}{0.0, 0.0, 0.55}
\definecolor{lightgray}{rgb}{0.929, 0.929, 0.929}

\newenvironment{formal}{%
  \MakeFramed{\advance\hsize-\width\FrameRestore}%
  \noindent\hspace{-4.55pt}
  \begin{adjustwidth}{}{7pt}%
  \vspace{2pt}\vspace{2pt}%
}
{%
  \vspace{2pt}\end{adjustwidth}\endMakeFramed%
}
\usepackage[compact]{titlesec}

\usepackage{times}
\usepackage{latexsym}

\usepackage[T1]{fontenc}

\usepackage[utf8]{inputenc}

\usepackage{microtype}

\usepackage{inconsolata}

\usepackage{graphicx}

\title{Instructions for *ACL Proceedings}

\author{Farnoosh Javadi, Phanideep Gampa, Alyssa Woo, Xingxing Geng 
 \\\bf Hang Zhang, \bf Jose Sepulveda, \bf Belhassen Bayar, \bf Fei Wang }

\title{LLM-based Weak Supervision Framework for Query Intent Classification in Video Search}

\begin{document}
\maketitle

\begin{abstract}

Streaming services have reshaped how we discover and engage with digital entertainment. Despite these advancements, effectively understanding the wide spectrum of user search queries continues to pose a significant challenge. An accurate query understanding system that can handle a variety of entities that represent different user intents is essential for delivering an enhanced user experience. We can build such a system by training a natural language understanding (NLU) model; however, obtaining high-quality labeled training data in this specialized domain is a substantial obstacle. Manual annotation is costly and impractical for capturing users' vast vocabulary variations. To address this, we introduce a novel approach that leverages large language models (LLMs) through weak supervision to automatically annotate a vast collection of user search queries. 
Using prompt engineering and a diverse set of LLM personas, we generate training data that matches human annotator expectations. By incorporating domain knowledge via Chain of Thought and In-Context Learning, our approach leverages the labeled data to train low-latency models optimized for real-time inference. Extensive evaluations demonstrated that our approach outperformed the baseline with an average relative gain of 113\% in recall. Furthermore, our novel prompt engineering framework yields higher quality LLM-generated data to be used for weak supervision; we observed 47.60\% improvement over baseline in agreement rate between LLM predictions and human annotations with respect to F1 score, weighted according to the distribution of occurrences of the search queries. Our persona selection routing mechanism further adds an additional 3.67\% increase in weighted F1 score on top of our novel prompt engineering framework.

\end{abstract}

\section{Introduction}

Media content discovery platforms have transformed how we explore and consume entertainment in the digital age~\cite{behare2024streaming,zhou2010impact}. However, comprehending and categorizing the diverse range of user search behaviors remains a significant challenge~\cite{liu2023improving}. Customers navigate vast content libraries using queries spanning various intents, from genre-specific searches like "\textit{comedy movies}" to queries based on cast and crew such as "\textit{Tom Hanks movies}", or thematic elements like "\textit{movies featuring talking animals}", etc~\cite{mansouri2008named}.

This diversity in search patterns underscores the necessity for a robust NLU system capable of recognizing and categorizing a wide range of entities\footnote{In this paper, we use intents and entities interchangeably for the sake of simplicity} – genres, cast and crew, themes, and many more. Such an entity recognition system is crucial for enhancing search functionality and delivering a better user experience. 

However, training a NLU model for query understanding requires lots of training data. Relying solely on implicit signals such as user click data for generating training data can be problematic~\cite{anand2023query}. Click data is inherently noisy, as user behaviors like clicks do not necessarily reflect the true intent behind a query. Direct labeling through human annotations can provide more accurate representations of query intent, but can be costly and impractical for covering all vocabulary variations at scale. To overcome these limitations, we propose a weak supervision approach that leverages the knowledge and capabilities of LLMs as a scalable proxy for direct labeling of query entities.

We further extend the use of LLMs by creating a network of LLM personas to assume different roles when interpreting a user's query intent, and aggregating their responses to form a collective output. This network of multiple personas essentially acts as an ensemble of LLMs, which can lead to better results compared to leveraging a single model alone. This persona-based approach is beneficial because accurately understanding queries in the movie domain requires strong knowledge of diverse movie vocabulary and its subdomains. While LLMs are trained on multiple domains, we need to use the personas in our prompting to systematically guide the LLM's parametric memory to concentrate on movie vocabulary. Furthermore, we propose a novel approach to select the best personas to respond for a given query by training a persona selection router model.

LLMs excel in natural language processing by capturing contextual nuances, making them ideal for query understanding. However, instead of relying on LLMs for real-time inference, which can be limited by higher latency and computational overhead, we leverage them to generate high-quality training data through weak supervision and employ computationally-efficient classification models for entity recognition during runtime. To the best of our knowledge, this is the first work that extensively studies query entity recognition for video search by leveraging LLMs through a weak supervision approach. Our key contributions are as follows: 


\begin{enumerate}
    \item We present a \textbf{novel prompt engineering framework} that combines In-Context Learning with intent-level examples, Chain of Thought with intent-level instructions and confidence level of predictions.  With this framework, we observe 47.60\% relative gain in terms of F1 score over baseline.
    
    \item Additionally, through a \textbf{multi-agent personas driven ensemble approach}, we can boost agreement rate between the LLM predictions and human annotations with respect to weighted F1 score by 3.67\% using a combination of personas in a multi-agent network compared to a no-persona single-agent using the novel prompt engineering framework. We propose a \textbf{persona routing mechanism} to select the best personas to predict the entities associated with a given query.
    
    \item Using the results from LLMs as training data for weak supervision, we propose  to train a \textbf{smaller low-latency model for real-time inference}, addressing the latency overhead associated with LLMs. Through extensive empirical evaluations, we demonstrate 113\% improvement in recall in query understanding and entity recognition compared to traditional NLU systems. 
\end{enumerate}

\label{section:introduction}

\section{Related Work}
Search query intent classification is an important step to drive the downstream retrieval process. Traditionally, intent classification has been solved through supervised classifiers~\cite{hashemi2016query, mendoza2009identifying,hu2009understanding, singhsurvey} and unsupervised techniques like clustering~\cite{cheung2012sequence, radlinski2010inferring, ren2014heterogeneous}. 

Leveraging the powerful reasoning and understanding capabilities of LLMs, researchers have used LLMs for query reformulation and expansion~\cite{wang2023query2doc, zhu2023large, anand2023query}. Recent studies have used LLMs like mT5 for query intent labeling via retrieval augmented generation (RAG) ~\cite{srinivasan2022quill}. However, these works employed relatively smaller LLMs and focused on datasets with limited intents, typically up to 5. Our approach utilizes much larger LLMs as teachers and incorporates advanced prompt engineering to handle 20+ intents.


Traditionally, various taxonomies with multiple levels have been developed across different domains for better query understanding. These datasets were labelled using logged user clicks \cite{lee2005automatic, singhsurvey, brenes2009survey}; however, clicks are subject to position and trust biases, and are only partial feedback from users, thus making them unreliable to label at scale~\cite{joachims2017unbiased, wang2016learning, zheng2021disentangling}. In our work, instead of click signals, we leverage powerful models like ClaudeV3 to label the data with high accuracy across multiple intents in the video domain and training a lightweight BERT model~\cite{devlin2018bert} for real time intent classification. 


LLMs reasoning capabilities are enhanced when a complex ambiguous problem is decomposed into multiple intermediate steps allowing it to perform a Chain of Thought (CoT) reasoning for solving the problem~\cite{wei2022chain}. Additionally, LLMs generalise well when provided with example demonstrations in the prompt allowing the model to learn from the examples following the In-Context Learning (ICL) paradigm~\cite{dong2022survey}. In our work, we designed a prompt that combines both CoT and ICL techniques. Furthermore, applying a persona to an LLM prompt to create distinctive output has been done in the context of generating diverse synthetic data~\cite{chan2024scalingsyntheticdatacreation}. We leverage this concept of using personas to generate diverse responses to identify entities associated with a given search query that are representative of the general population. We further build on this concept by proposing our novel routing mechanism for persona selection for a given search query.

\label{section:Related Work}

\section{Approach}
\label{section:approach}


Training machine learning models for query entity classification requires extensive annotated data covering a wide range of vocabulary and intents. Manual annotation by humans is costly and impractical for comprehensive coverage. Using LLMs for annotation is a more cost-effective solution that can handle diverse vocabulary comprehensively. Building upon ICL and CoT prompting techniques, we leverage a network of persona based LLMs to generate training data by annotating a sizable collection of customer search queries. To overcome LLM latency challenges, we train a student BERT model on this generated training data also enabling real-time inference.  


\textbf{Query Entities:} When exploring content on streaming services, customers use a diverse range of search queries. They may search by genre, such as "\textit{comedy movies}", or by audio language, like "\textit{french movies}", or by year of release, for instance, "\textit{2023 TV show}", among many other criteria. We compiled a set of more than 20 entities that capture customer intent in their search. We provided the most important entities, their explanations and examples in Table \ref{tab:entity_table}.

\subsection{Prompt Engineering Techniques}


The LLMs are prompted with instructions on identifying and tagging entities within the queries. Extensive prompt engineering techniques are employed to enhance the effectiveness of these prompts.

 \textbf{Chain of Thought (CoT):} Through CoT prompting~\cite{wei2022chain}, we guide the model through the reasoning process required to categorize entities into predefined classes. This method can enhance the model's accuracy and understanding by breaking down the classification task into a series of logical steps. We first define each entity clearly. We then conduct a thorough examination against each definition iteratively, requiring the model to report all intermediate findings, as shown in Figure \ref{figure:COT Steps}.

\begin{figure}[H]

  \begin{formal}
      \textbf{CoT Steps}:\\
      \textbf{\textit{Step1}}: Compare the query against each entity category iteratively. Use the provided examples as reference in making the decisions.\\
      \textbf{\textit{Step2}}: Check if the query is a IntentMovie entity. A query is a IntentMovie entity if it focuses on the user's intent to find movies, whether through the mention of a a specific description, \
characteristic, actor, or genre.\\
      \textbf{\textit{Step3}}: Check if the query contains the CastAndCrew entity. A query is a CastAndCrew entity if it contains the name of an actor, director, etc. \\
    \textbf{\textit{Step4}}: ... \\
    \textbf{\textit{StepK}}: Assign the label \textit{None} to the query if it does not fit into any of the specified entity categories mentioned above. \\
  \end{formal}
  \caption{Demonstrative COT Steps in LLM prompt.}
      \label{figure:COT Steps}
  \vspace{-10pt}
  \end{figure}
    
 \textbf{In Context Learning (ICL):}
    Another key technique employed in our prompt design is In-Context Learning (ICL)~\cite{dong2022survey}, as certain entities may require domain-specific knowledge — for instance, StreamingService, AudioLanguage, and others. To equip the LLM with the necessary context to accurately annotate these specialized entities, we append representative relevant examples of each entity to the prompt, as demonstrated in Figure \ref{figure:ICL steps}. By providing these in-context examples upfront, we aim to enhance the LLM's understanding of the domain knowledge required for proper annotation of such entities. 

\begin{figure}[H]

  \begin{formal}
      \textbf{ICL Instructions}:\\
      \textbf{\textit{AudioLanguage} entity examples:} Arabic, Bangla, Chinese, English, etc.\\
      \textbf{\textit{Holiday} entity examples:} Christmas, Thanksgiving, Easter, etc.\\
      \textbf{\textit{StreamingService} entity examples:} HBO Max, Amazon Prime, Apple TV, Netflix,etc.\\
      \textbf{\textit{Sport} entity examples:} football, Manchester United, Nadal, etc.\\

  \end{formal}
  \caption{Sample ICL instructions in LLM prompt.}
      \label{figure:ICL steps}
  \vspace{-10pt}
  \end{figure}
    
  \textbf{Adding Confidence Score:} We ask LLMs to provide a confidence level for each entity — categorized as high, medium, or low. 
    LLMs are powerful but can sometimes make mistakes or generate outputs with varying levels of certainty. These scores quantify the model's certainty in its predictions. High-confidence outputs are processed automatically or used for decision-making, while medium and low-confidence outputs undergo human review or additional validation. This approach mitigates risks associated with uncertain predictions.

\subsection{Personas}

\begin{figure}
    \centering
    \includegraphics[scale=0.23]{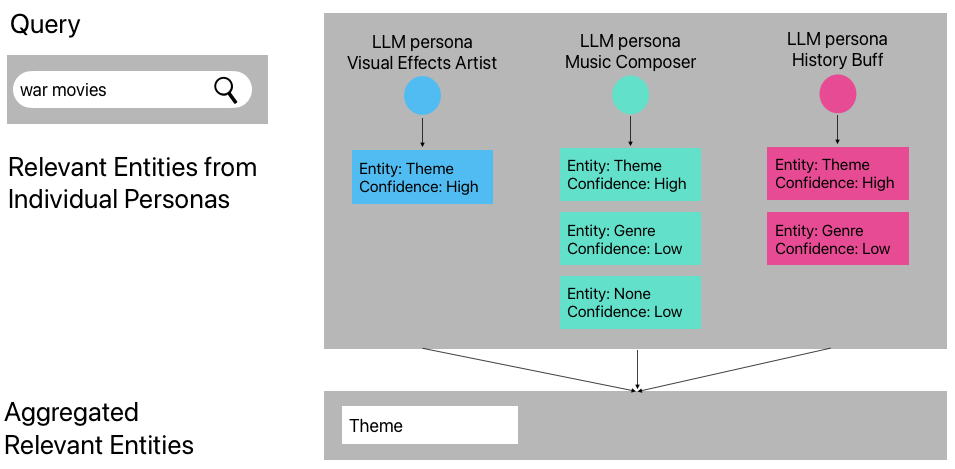}
    \caption{Multi-agent LLM persona network}
    \label{fig:persona_response_aggregation}
\end{figure}

We performed experiments using a multi-agent LLM persona network, representing human annotators with different experiences and personalities. A persona is a personality or role that we assign to the LLM to play when responding to a prompt, each with a specialized knowledge and perspective. Each one of these decentralized personas come together to solve problems that may require several diverse views, such as in our application, as understanding user query intent can be an ambiguous task. We create a repository of diverse personas by using our domain knowledge to manually craft some personas to begin, then synthesizing the remaining personas using ChatGPT3. Representative personas and their descriptions used in experimentation are in \ref{tab:personas}. Figure \ref{fig:persona_response_aggregation} displays a visualization of the multi-agent LLM persona network and aggregation for the search query understanding task.

\begin{figure}
    \centering
    \includegraphics[scale=0.25]{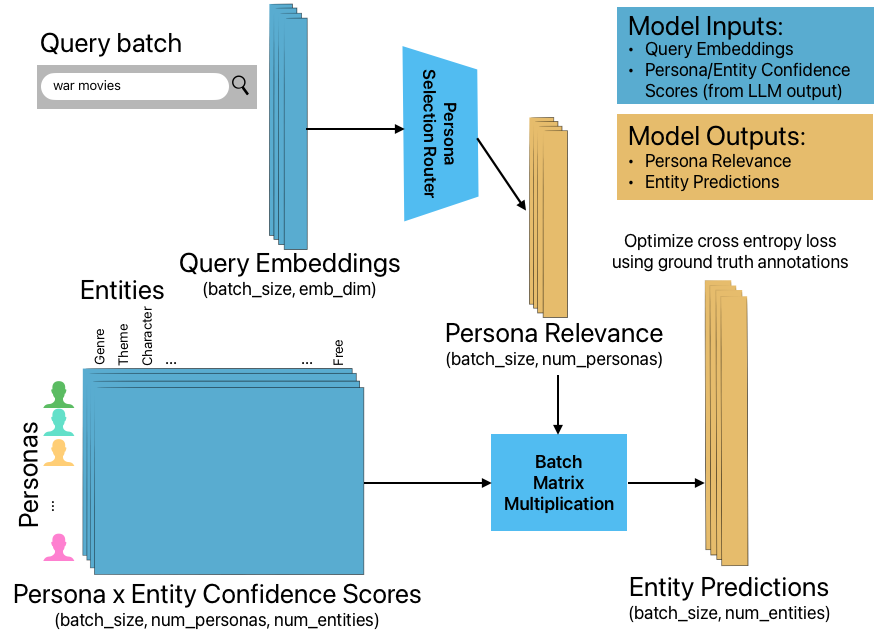}
    \caption{Persona Selection Mechanism via Router}
    \label{fig:persona_router}
\end{figure}

Additionally, because we have a large repository of diverse personas, and each persona has different expertise, we need a method of strategically selecting which personas to use to perform a task given a query as oppose to selecting randomly. We propose a persona routing mechanism, as summarized in Figure \ref{fig:persona_router}. The model takes in two inputs: (1) semantic embedding to represent queries using Cohere model~\footnote{https://docs.cohere.com/docs/models\#embed}, and (2) Persona $\times$ Entity Confidence Score matrices, which represent the LLM responses for a set of queries for every persona in the repository. The Persona $\times$ Entity Confidence Score matrices are formed by converting the ``Low'', ``Medium'', ``High'' confidence scores to numeric values 1-3, respectively, and taking 0 for any entity the LLM did not select as relevant to the query.

The primary task is to learn a ``Persona Selection Router", which consists of two linear layers, a dropout layer, and a ReLU activation, which produces an output that represents the relevance of each persona for a given query. Because we lack a ground truth to learn a persona's relevance for a given query, we train an auxiliary task to predict entities for a given query, for which we have annotation data to optimize on. We do so by taking the Persona $\times$ Entity Confidence Score matrix input and performing a batch matrix multiplication with the persona relevance to obtain the entity prediction output. We optimize a cross-entropy loss on the entity prediction output during model training to ultimately train our Persona Selection Router to produce accurate persona relevance output.

\subsection{Multi-Label Entity Classifier}
 \label{section: multi label model}
\begin{figure}
\centering
\includegraphics[width=3 in]{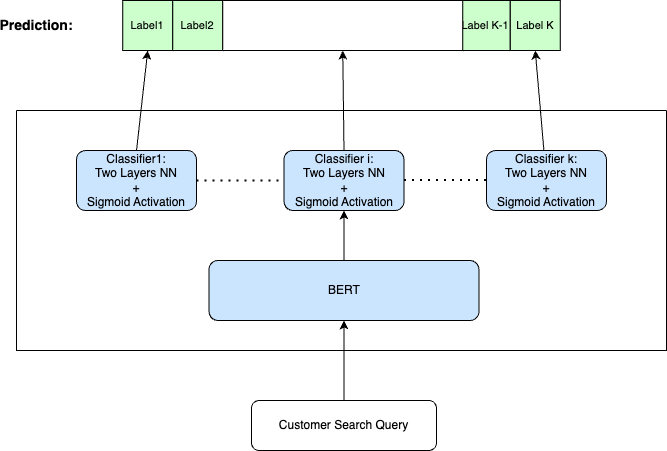}
\caption{NLU Model Structure}
\label{fig:structure}
\end{figure}

To interpret and classify search queries based on predefined entities during run-time, we propose training a multi-label entity classifier built upon the BERT (Bidirectional Encoder Representations from Transformers) model~\cite{devlin2018bert}. BERT is a powerful, pre-trained language model that can be fine-tuned for various natural language processing tasks. Its bidirectional contextual representation and self-attention mechanism allow for a deep understanding of context and semantics within short text sequences, making it well-suited for classifying search queries. Moreover, the BERT model offers low latency, which is essential for real-time query entity classification in production environments. The classifier architecture is shown in Figure ~\ref{fig:structure}.

\section{Experiments}

\subsection{LLM Prompt Design}
The aim of this experiment is to benchmark different LLMs and prompt engineering techniques to find the best one to generate LLM data to be used for weak supervision, that best aligns with human expectations. We benchmark the performance of LLM-generated annotations against manual annotations by human experts to assess the viability of using LLMs as a scalable and cost-effective alternative for annotation processes.
\subsubsection{Data}
For manual annotation, we selected a sample of 800 search queries from the first two weeks of April 2024. The queries were sampled weighted by the frequency of searches during that period.
\subsubsection{Experiment Setup}
 We evaluate and compare several approaches for entity recognition and annotation. First, we benchmark the performance of Claude V2.1, Claude V3 Sonnet, Claude V3 Haiku, Mixtral - 8x7B moE , Mistral 7b LoRA~\cite{DBLP:journals/corr/abs-2106-09685} fine-tuned model. Additionally, we assess the effectiveness of different prompt engineering strategies, including a baseline prompt, a prompt incorporating confidence level, CoT, and ICL techniques. In addition we present the benefit of personas. We present both unweighted (standard) and weighted metrics, which take into account the frequency distribution of search queries.

\subsubsection{Results}
\begin{table*}[!ht]
\caption{
Ablation Study for benchmarking across different prompt engineering techniques. We report the percentage improvement over the Baseline prompt.}
\vspace{2mm}
\label{tab:prompt_benchmark}
\centering
\resizebox{6.5in}{!}{%
{\footnotesize
\begin{tabular}{@{}cccccccccccc@{}}\toprule
& \multicolumn{3}{c }{\makecell{{Baseline + Confidence Level}}} & \phantom{ab}& \multicolumn{3}{c}{\makecell{{Baseline + Confidence Level + CoT}}} & \phantom{ab}& \multicolumn{3}{c}{\makecell{{Baseline + Confidence Level + CoT + ICL}}}\\
\cmidrule{2-4} \cmidrule{6-8} \cmidrule{10-12} 
& \makecell{F1} & \makecell{Precision} & \makecell{Recall} & & \makecell{F1} & \makecell{Precision} & \makecell{Recall} && \makecell{F1} & \makecell{Precision} & \makecell{Recall} \\  
(Unweighted) \\
\texttt{Claude-v3 Sonnet} & $0.90\%$ & $2.91\%$ & $-0.009\%$ && $2.76\%$ & $1.85\%$ & $4.60\%$ &&  $\textbf{4.06\%}$ & $4.38\%$ & $3.76\%$\\ \\
(Weighted) \\
\texttt{Claude-v3 Sonnet} & $4.88\%$ & $3.81\%$ & $-0.01\%$ && $05.05\%$ & $9.30\%$ & $7.49\%$  && $\textbf{6.70\%}$ & $7.29\%$ & $6.15\%$ \\ 
\bottomrule
\end{tabular}
}
}
\end{table*}

\begin{table*}[!ht]
\caption{
Benchmarking across different LLMs on manual annotated data. Weighted metrics take into account the frequency distribution of search queries. We picked the prompt incorporating Confidence Level + CoT + ICL. We report the percentage improvement over the Baseline (Lexical Matching).}
\vspace{2mm}
\label{tab:LLM_benchmark}
\centering
\resizebox{5.6in}{!}{%
{ \footnotesize
\begin{tabular}{@{}cccccccccccc@{}}\toprule
& \multicolumn{3}{c}{\makecell{{Unweighted}}} & \phantom{ab}& \multicolumn{3}{c}{\makecell{{Weighted}}} \\
\cmidrule{2-4} \cmidrule{6-8} 
& \makecell{F1} & \makecell{Precision} & \makecell{Recall} & & \makecell{F1} & \makecell{Precision} & \makecell{Recall} \\  
\texttt{Claude-v3 Sonnet} & $\textbf{54.45\%}$ & $\textbf{18.41\%}$ & $\textbf{89.89\%}$ && $\textbf{47.60\%}$ & $\textbf{9.56\%}$ & $\textbf{83.74\%}$\\
\texttt{Claude-v3 Haiku } & $48.81\%$ & $11.32\%$ & $87.39\%$ && $44.99\%$ & $10.09\%$ & $76.73\%$ \\ 
\texttt{Claude-v2.1} & $49.63\%$ & $17.72\%$ & $79.59\%$ && $43.12\%$ & $-1.42\%$ & $81.34\%$ \\
\texttt{Mixtral - 8x7B MoE } & $2.81\%$ & $11.98\%$ & $44.05\%$ && $25.93\%$ & $-2.50\%$ & $55.13\%$ \\ 
\texttt{Mistral 7B LoRA finetuned } & $5.85\%$ & $-10.79\%$ & $9.99\%$ && $1.46\%$ & $0.73\%$ & $0.51\%$ \\ 

\bottomrule
\end{tabular}
}
}
\end{table*}

The results presented in Table ~\ref{tab:LLM_benchmark} demonstrate that the Claude-3 Sonnet model has the best overall performance. Additionally, the results presented in Table ~\ref{tab:prompt_benchmark} indicate that adding confidence level, CoT and ICL techniques enhance the effectiveness of the prompts, leading to improved query understanding and entity recognition capabilities. When adding personas compared to the no persona baseline, we see 0.81\% increase in weighted F1 score using random 3 personas, and 3.67\% increase using the best 3 personas via the router (Table \ref{tab:persona_test_eval}). This experiment was conducted on a test set composed of 20\% of entity-rebalanced annotation data, while the persona selection router model was trained using 80\% of entity-rebalanced annotated data.

\begin{table}[!ht]
\caption{
Test set gains of random persona selection and persona selection via router compared to no persona (Baseline + Confidence Level + CoT + ICL from Table~\ref{tab:prompt_benchmark}).}
\vspace{2mm}
\label{tab:persona_test_eval}
\centering
\resizebox{3.0in}{!}{%
{\footnotesize
\begin{tabular}{@{}cccccccccccc@{}}\toprule
& \multicolumn{3}{c }{\makecell{{Random 3 Personas}}} & \phantom{ab}& \multicolumn{3}{c}{\makecell{{Best 3 Personas via Router}}} \\
\cmidrule{2-4} \cmidrule{6-8}
& \makecell{F1} & \makecell{Precision} & \makecell{Recall} & & \makecell{F1} & \makecell{Precision} & \makecell{Recall}\\  
(Unweighted) \\
\texttt{Claude-v3 Sonnet} & $-4.35\%$ & $-3.36\%$ & $-5.49\%$ && $\textbf{3.27\%}$ & $5.04\%$ & $1.24\%$\\ \\
(Weighted) \\
\texttt{Claude-v3 Sonnet} & $0.81\%$ & $1.37\%$ & $0.20\%$ && $\textbf{3.67\%}$ & $5.67\%$ & $1.51\%$ \\ 
\bottomrule
\end{tabular}
}
}
\end{table}


\subsection{Entity Classifier Experiments}

The objective of this experiment is to conduct a performance evaluation and comparative analysis among three entity recognition systems:

\begin{itemize}
    \item A baseline approach leveraging lexical matching to identify and label entities within text.
    \item A multi-label entity classifier trained on labels derived from the lexical matching baseline.
    \item A multi-label entity classifier trained on labels generated by Claude V3 Sonnet.
\end{itemize}
\subsubsection{Data}

We randomly sampled 100,000 user search queries from the top one million popular queries. This dataset was split into 20\% for testing, 70\% for training and 10\% for dev set. We employed Claude V3 to annotate all queries with their relevant entities from a universe of 22 entities. The Claude V3 annotations on the training set were used to train the multi-label entity classifier, while the annotations on the test set served as the reference ground truth for evaluating the performance of our proposed multi-label entity classifier and the baseline lexical matching approach. The dev set was used for hyperparameter tuning.

\subsubsection{Experiment Setup}
Our approach for building multi-label entity classifier leverages the powerful pre-trained BERT-base model. We extract the contextualized representations  from the last four hidden layers of the BERT model and concatenate them to form a comprehensive sequence representation. This sequence representation is then fed into entity-specific classifiers, consisting of two layers of neural networks followed by a Sigmoid activation function. These classifiers independently generate probabilities for 22 entities. For training the entity classifier, we employed the binary cross-entropy (BCE) logistic loss function. The model was optimized using the AdamW optimizer ~\cite{loshchilov2019decoupled} with a learning rate of $10^{-5}$. 



\subsubsection{Results}
We evaluate the model's performance using precision and recall metrics because we are comparing against a lexical matching baseline. For each entity, we picked the thresholds that match the lexical baseline's recall and precision, respectively. Table \ref{tab:po_kto_dpo} presents the overall precision and recall scores for the models. Refer to ~\ref{tab_appendix:po_kto_dpo} for the performance on most important entities. Also ~\ref{tab:ground_truth_MLC} shows substantial improvement on the 800 manually annotated data.

\begin{table}[h]
\caption{ Comparison between multi-label entity classifier trained on labels generated by Claude V3 and one trained on labels derived from the baseline lexical matching. We report the percentage
improvement over the baseline which is lexical matching only.}
\vspace{2mm}
\label{tab:po_kto_dpo}
\centering
\resizebox{3.0in}{!}{%
{\footnotesize
\begin{tabular}{@{}cccc@{}}
\toprule

& \makecell{Entity Classifier\\lexical matching labels} & \makecell{Entity Classifier\\Claude labels} \\
\cmidrule(lr){2-3}
\texttt{Precision@Matching Recall}& -12\% &35\% \\
\texttt{Recall@Matching Precision} & 83\% & 113\% \\
\bottomrule
\end{tabular}
}
}
\end{table}




\section{Conclusion}
This paper presented a novel approach leveraging LLMs and prompt engineering techniques, including multi-persona selection to enhance query entity recognition in the video domain. Our method addresses limitations of relying solely on user click data, lexical matches and manual annotations for training data generation. By combining domain knowledge with LLMs' contextual understanding capabilities, our framework accurately recognizes a wide range of entities in user queries, enabling video platforms to deliver more relevant search results. To mitigate LLMs' latency challenges during real-time inference, we proposed a proxy solution using efficient weakly supervised classification models trained on LLM-generated data, ensuring low-latency performance. Through extensive evaluations, our approach outperformed traditional NLU systems in accurately identifying query entities, demonstrating a substantial 113\% relative gain in recall. Our prompt engineering framework for LLM data generation achieved a 47.60\% improvement in weighted F1 score agreement between LLM predictions and human annotations, compared to a lexical matching baseline. Additionally, our persona selection router mechanism provided a further 3.67\% boost.


\clearpage

\bibliography{acl_latex.bib}

	


\onecolumn
\begin{appendices}
\setcounter{table}{0}
\renewcommand{\thetable}{A\arabic{table}}
\renewcommand\thefigure{\thesection.\arabic{figure}} 


\section{Most important entities that represent user search query intents}
\begin{table*}[h]
    \centering
    \caption{Query Intent Understanding Entities}
    \begin{tabular}{|l|p{8cm}|p{5cm}|}
        \hline
        \textbf{Entities} & \textbf{Definition} & \textbf{Examples} \\
        \hline
        \textbf{IntentMovie} & Focuses on the user's intent to find movies, whether through the mention of a a specific description, characteristic, actor, director, or genres.  & Leonardo DiCaprio movies, Tarantino movies, Spider man movies, movies like Interstellar \\
        \hline
        \textbf{IntentTvSeries} & Focuses on the user's intent to find TV shows or TV series that fit a specific description, characteristic, actor, director, or genre. & recent shows, best series, drama TV series, TV series like The Marvelous Mrs. Maisel \\
        \hline
        \textbf{Theme} & Refers to a concept describing the details of a content-type. & after dark, alien abduction, alligator, based on true story \\
        \hline
        \textbf{Genre} & Represents the genre of video media content;  & horror movies, romantic comedy, action-comedy movies\\
        \hline
        \textbf{CastAndCrew} & Refers to actors, directors, writers, authors or any other famous personalities who are related to the entertainment industry or are subjects of media contents & Tom Hanks movies \\
        \hline
        \textbf{TVSeriesName} & A query is a TVSeriesName entity if it searches for a specific TV series/season title. & The Marvelous Mrs. Maisel \\
        \hline
        \textbf{MovieName} & A query is a MovieName entity if it searches for a specific movie title. & Interstellar, Oppenheimer \\
        \hline
        \textbf{StreamingService} & Is an entity or a company in the entertainment industry which offers exclusive videos on their platform.  & HBO Max, Amazon Prime, Apple TV, Netflix \\
        \hline
        \textbf{Recency} & Refers to queries for recent movies, videos, etc. & newly-released movies, new movies, recent TV shows \\
        \hline
        \textbf{Popularity} & Refers to queries for popular movies/videos. & popular movies, widely-viewed movies, highly-streamed movies \\
        \hline
        \textbf{ReleaseYear} & Is a year in which titles are released. & 2023 movies \\
        \hline
        \textbf{Decade} &  Refers to queries where users specify a particular decade in which video was released. & 90's movies, 80s shows, 1990 movies\\
        \hline
        \textbf{FreeContent} & Refers to queries searching for free movies/videos. & free shows, free movies \\
        \hline
        \textbf{AudioLanguage} & Represents the language of video media content & Arabic movies, Bangla movies \\
        \hline
        \textbf{Franchise} & Refers to a series of related movies or videos marketed under a single brand name. & Avengers, Batman \\
        \hline
        \textbf{Holiday} & Refers to queries where users are looking for movies or content related to specific holidays or celebrations. & Christmas movies, Thanksgiving \\
        \hline
        \textbf{Sport} & Refers to queries related to sports events, teams, athletes, or sports-related content. & football, Manchester United, Serena Williams\\
        \hline
        \textbf{Character} & Is a figure (i.e. person, animal or inanimate object) that drives the story forward either in a movie, or a TvSeries, or a franchise or across all of the three. & Asterix, Batman, Blippi, Charlie Brown\\
        \hline 
    \end{tabular}
    \label{tab:entity_table}
\end{table*}




\clearpage
\setcounter{table}{0}
\renewcommand{\thetable}{\arabic{table}}
\renewcommand\thefigure{\thesection.\arabic{figure}} 

\section{Entity Level Performance of the Entity Classifier}




\begin{table*}[h]
\caption{ Entity level performance Comparison Between multi-label entity classifier trained on labels generated by Claude V3, Multi-Label Entity Classifier trained on labels derived from the lexical matching and Baseline(lexical matching). We report the percentage improvement over the Baseline(lexical Matching)}
\vspace{2mm}
\label{tab_appendix:po_kto_dpo}
\centering
\resizebox{6.2in}{!}{%
{\footnotesize
\begin{tabular}{@{}cccccccccccc@{}}\toprule
& \multicolumn{2}{c}{\makecell{{Precision@Matching Recall}}} & \phantom{ab}& \multicolumn{2}{c}{\makecell{{Recall@Matching Precision}}}\\
\cmidrule(lr){2-3}
\cmidrule(lr){5-6}
& \makecell{Entity Classifier (Claude labels)} & \makecell{Entity Classifier (Lexical labels)}  & & \makecell{Entity Classifier (Claude labels)} & \makecell{Entity Classifier (Lexical labels)} \\  

\texttt{Theme} & $71.42\%$ & $-35.71\%$ && $1775\%$ & $1700\%$ \\
\texttt{Genre} & $22.5\%$ & $-0.45\%$ && $92\%$ & $91\%$  \\
\texttt{CastAndCrew} & $28.94\%$ & $-7.89\%$ && $100\%$ & $56.81\%$ \\
\texttt{TvSeriesName} & $262.96\%$ & $3.70\%$ && $280\%$ & $88.46\%$ \\
\texttt{MovieName} & $68.42\%$ & $-19.29\%$ && $100\%$ & $102.04\%$ \\
\texttt{Holiday} & $2.04\%$ & $-7.14\%$ && $54.45\%$ & $50\%$ \\
\texttt{ReleaseYear} & $8.79\%$ & $1.09\%$ && $150\%$ & $153.57\%$ \\
\texttt{AudioLanguage} & $5.37\%$ & $-5.37\%$ && $40.74\%$ & $33.33\%$ \\
\bottomrule
\end{tabular}}
}
\end{table*}

\begin{table*}[!ht]
\caption{
Percentage improvement of a multi-label entity classifier trained on labels generated by Claude V3 over the baseline (lexical matching). The entity classifier achieves results comparable to those of Claude V3.}
\vspace{2mm}
\label{tab:ground_truth_MLC}
\centering
\resizebox{3.0in}{!}{%
{\footnotesize
\begin{tabular}{@{}lccc@{}}\toprule
& F1 & Precision & Recall \\
\midrule
\texttt{Unweighted} & 51.07\% & 13.49\% & 92.97\% \\
\texttt{Weighted} & 45.54\% & 7.14\% & 84.61\% \\
\bottomrule
\end{tabular}
}
}
\end{table*}

\clearpage
\setcounter{table}{0}
\renewcommand{\thetable}{C\arabic{table}}
\renewcommand\thefigure{\thesection.\arabic{figure}} 
\section{Personas for evaluating query intent understanding task}
We used our domain knowledge to create sample personas that may be knowledgeable about streaming services and search queries in the context of media streaming. We provided the sample personas to \textsc{ChatGPT3} and asked it to generate additional personas. We used our domain knowledge to categorize the crafted and generated personas into categories: experts, non-domain experts, and niche experts. Table \ref{tab:personas} shows the details of few personas out of 32 we experimented with. Personas with a $^*$ are human-crafted.

\begin{table}[h]
    \caption{Sample personas used in evaluation of video title similarity task}
    \centering
    \begin{tabular}{|l|p{1.2in}|p{3.3in}|}
    \hline
    Category & Persona Name & Persona Description\\
    \hline
    Experts & \textsc{Merchandiser}$^*$ & You are a merchandiser for a popular online streaming service and have extensive knowledge of the movie and TV show catalog. You are very knowledgeable about crafting recommendations for a wide range of audiences that cater to their different preferences. \\\cline{2-3}
         & \textsc{Movie Critic}$^*$ & You are a famous movie and tv critic. Your friend has asked you for honest advice about titles that are similar to one that they just watched. You want to help them sincerely. Make sure that the recommendations are good! You like your friend. \\\cline{2-3}
         & \textsc{Movie Buff}$^*$ & You are an avid movie enthusiast and are part of several movie clubs. As such, you are very good at recommending related movies and sorting out the not so good ones. \\
         \hline
    Non-Domain Experts 
         & \textsc{Book Club Member} & You're a 55-year-old avid reader and member of a local book club. While you love diving into novels, sometimes you enjoy switching things up by watching book adaptations or exploring documentaries related to literature on streaming platforms. \\\hline
    Niche Experts
         & \textsc{Horror Aficionado} & You're a 35-year-old horror aficionado with a love for all things spooky and supernatural. Your streaming watchlist is packed with horror movies, suspenseful thrillers, and chilling TV series, and you eagerly anticipate the latest releases in the genre.\\
    \hline
    \end{tabular}
    \label{tab:personas}
\end{table}{}

\setcounter{table}{0}
\renewcommand{\thetable}{D\arabic{table}}
\renewcommand\thefigure{\thesection.\arabic{figure}}


\end{appendices}

\end{document}